\documentclass[11pt]{article}
\usepackage{blois,epsfig}

\def\Journal#1#2#3#4{{#1} {\bf #2}, #3 (#4)}

\def\be{\begin{equation}}
\def\ee{\end{equation}}
\def\bea{\begin{eqnarray}}
\def\eea{\end{eqnarray}}

\begin{document}
\vspace*{4cm}
\title{COMMENTS ON "MEASURING THE GRAVITY SPEED BY VLBI"} 

\author{ H. ASADA }

\address{Institute for Astrophysics at Paris, 
98bis boulevard Arago, 75014 Paris, France\\
Faculty of Science and Technology, Hirosaki University, 
Hirosaki 036-8561, Japan}

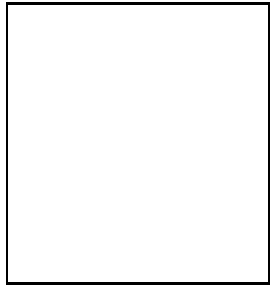
\maketitle\abstracts{
Einstein gravity with extra dimensions or alternative gravity theories 
might suggest that the gravity propagation speed can be different from 
the light speed. Such a difference may play a vital role in 
the primordial universe. In recent, Kopeikin and Fomalont claimed 
the first measurement of the gravity speed by VLBI. 
However, the measurement has no relevance with the speed of gravity 
as I had shown before the observation was done. 
It seems that our conclusion has been established well 
by re-examining recent papers with great care. 
}

\section{Introduction}
Einstein gravity with extra dimensions or alternative gravity theories 
might suggest that the gravity propagation speed can be different from 
the light speed. 
The Shapiro time delay plays an important role in experimental 
verification of Einstein's theory of general relativity. 
In fact, VLBI (very long baseline interferometry) confirms 
the validity of general relativity. 
The accuracy will be achieved within a few picoseconds (ps), 
namely about 10 microarcseconds $(\mu as)$, for instance 
by VERA (VLBI Exploration of Radio Astrometry \cite{HKS}). 
This is why the correction to the Shapiro time delay has been
intensively investigated \cite{KS,Kopeikin01}. 
Kopeikin found that the excess time delay caused by Jupiter 
can be measured on the 8th of September 2002. He also argued 
that the excess was due to the propagation of gravity, 
which could be tested through the observation. 
Indeed, Kopeikin and Fomalont made the observation and claimed 
the first measurement of the {\it gravity} speed \cite{FK}. 
Before the measurement, however, it had been shown that 
it comes from the propagation of light but not gravity \cite{Asada}. 

The primary reason against his conclusion is based on the
post-Newtonian approximation of general relativity: 
In the approximation, we perform expansions in the inverse of 
light velocity $c$, by considering all quantities are 
perturbations around the Newtonian parts.  
The deflection of light, the perihelion shift of Mercury and 
the time delay occur at the first post-Newtonian order $O(c^{-2})$. 
Actually, using these three effects, the classical tests 
have been done to confirm the validity of general relativity 
(For a thorough review, see Will \cite{Will}). 
The propagation of gravity appears at $O(c^{-4})$ because 
mass dipole moments vanish in contrast to electromagnetism 
\cite{Thorne,Will}. 
The effect of the radiation reaction of quadrupole gravitational 
waves \cite{PM,Thorne,Will} 
has been confirmed through the observation of decaying orbital 
period of Hulse-Taylor binary pulsar \cite{Taylor}. 
On the other hand, the Kopeikin's excess time delay 
for the standard Shapiro delay at $O(c^{-2})$ is $O(v_J/c^3)$, 
where $v_J$ is the velocity of Jupiter. The order of the excess 
is lower than that of the propagation of gravity, so that 
the excess cannot be caused by the gravity propagation. 
The origin of the excess is clarified in the following.

\section{Shapiro delay in retarded time}
Since a massive body produces gravitational fields as a curved 
spacetime, a light signal will take a longer time to traverse 
a given spatial distance than it would if Newtonian theory were valid. 
In deriving the Shapiro time delay, the Einstein equation for 
the gravitational field and the null geodesics for the light ray 
are solved up to $O(c^{-2})$. In particular, the Einstein equation 
is reduced to Poisson-type equations, so that the propagation 
of gravity is not incorporated. The Shapiro delay for 
a light signal from an emitter to an observer is obtained 
in a logarithmic form \cite{Shapiro,Will}. 

Let us consider a baseline denoted by $\mbox{\boldmath $B$}$; 
at $t_1$ and $t_2$, the light signals from a quasar reach 
the first and second stations which locate at 
$\mbox{\boldmath $x$}_1(t)$ and $\mbox{\boldmath $x$}_2(t)$, 
respectively, so that we can define the baseline as 
the spatial interval between the simultaneous events 
$\mbox{\boldmath $B$}=\mbox{\boldmath $x$}_1(t_1)
-\mbox{\boldmath $x$}_2(t_1)$. 
Each station is denoted by $i$ later. 
Since the Shapiro delay is a consequence of integration of 
the null geodesics on the light cones, it is convenient and 
crucial to use $s_1$ and $s_2$, retarded time which is constant 
on each light cone emanating from events 
$(t_1, \mbox{\boldmath $x$}_1(t_1))$ 
and $(t_2, \mbox{\boldmath $x$}_2(t_2))$, 
so that we can have 
$\mbox{\boldmath $x$}_i(t_i)=\mbox{\boldmath $x$}_i(s_i)$ 
for $i=1, 2$. 
Hence, the difference of the Shapiro delay between the baseline 
is expressed as \begin{equation}
\Delta(t_1,t_2)=\frac{2GM}{c^3}\ln
\frac{R_{1J}+\mbox{\boldmath $K$}\cdot\mbox{\boldmath $R$}_{1J}}
{R_{2J}+\mbox{\boldmath $K$}\cdot\mbox{\boldmath $R$}_{2J}} , 
\label{eq:shapiro}
\end{equation}
where the unit vector from the Earth to the emitter of the light 
is denoted by $\mbox{\boldmath $K$}$, the position of Jupiter 
by $\mbox{\boldmath $x$}_J(t)$ and we defined 
$\mbox{\boldmath $R$}_{iJ}=\mbox{\boldmath $x$}_i(s_i)
-\mbox{\boldmath $x$}_J(s_i)$ and 
$R_{iJ}=|\mbox{\boldmath $R$}_{iJ}|$ on each light cone 
labeled by $i=1, 2$. 
Equation ($\ref{eq:shapiro}$) is exact at the post-Newtonian order. 
In order to clarify the origin of the excess, we expand 
this equation approximately as follows. 

Since the speed of Jupiter $v_J$ is much {\it smaller} than $c$, 
we find 
\begin{eqnarray}
\mbox{\boldmath $R$}_{iJ}
&=&\mbox{\boldmath $r$}_{iJ}
+\frac{\mbox{\boldmath $v$}_J}{c}r_{iJ}+O(c^{-2}) , \\
R_{iJ}&=&r_{iJ}+\frac{\mbox{\boldmath $R$}_{iJ}
\cdot\mbox{\boldmath $v$}_J}{c}+O(c^{-2}) , 
\end{eqnarray}
where we used 
$\mbox{\boldmath $x$}_i(s_i)=\mbox{\boldmath $x$}_i(t_i)$ 
and denoted the spatial displacement vector between 
the simultaneous events by 
$
\mbox{\boldmath $r$}_{iJ}=\mbox{\boldmath $x$}_i(t_i)
-\mbox{\boldmath $x$}_J(t_i), 
$
and the interval by $r_{iJ}=|\mbox{\boldmath $r$}_{iJ}|$. 
Furthermore, since $B$ is much {\it shorter} than 
$R \sim R_{iJ}$, we obtain 
\begin{equation}
r_{2J}-r_{1J}=\mbox{\boldmath $N$}_{1J}\cdot\mbox{\boldmath $B$}
+O\left(\frac{B^2}{R}\right) , 
\end{equation}
where we defined 
$\mbox{\boldmath $N$}_{1J}=\mbox{\boldmath $r$}_{1J}/r_{1J}$. 
Hence, Eq. $(\ref{eq:shapiro})$ becomes 
\begin{eqnarray}
\Delta(t_1,t_2)=\frac{2GM}{c^3}&\Bigl(&
\ln
\frac{r_{1J}+\mbox{\boldmath $K$}\cdot\mbox{\boldmath $r$}_{1J}}
{r_{2J}+\mbox{\boldmath $K$}\cdot\mbox{\boldmath $r$}_{2J}} 
\nonumber\\
&&-\frac{\mbox{\boldmath $B$}\cdot\mbox{\boldmath $v$}_J
+(\mbox{\boldmath $N$}_{1J}\cdot\mbox{\boldmath $B$})
(\mbox{\boldmath $K$}\cdot\mbox{\boldmath $v$}_J)}
{c(r_{1J}+\mbox{\boldmath $K$}\cdot\mbox{\boldmath $r$}_{1J})}
+O(c^{-2}) \Bigr) . 
\label{eq:shapiro2}
\end{eqnarray}

We denote by $\theta$ a {\it small} angle between 
the first station - source and the station - Jupiter 
at simultaneous time $t$. 
We obtain 
\begin{eqnarray}
\mbox{\boldmath $N$}_{1J}&=&-\mbox{\boldmath $K$}\cos\theta
+\mbox{\boldmath $n$}\sin\theta \nonumber\\
&=&-\left(1-\frac{\theta^2}{2}\right)\mbox{\boldmath $K$}
+\theta\mbox{\boldmath $n$}+O(\theta^3) , 
\end{eqnarray}
where $\mbox{\boldmath $n$}$ is a unit normal vector from the Jupiter 
to the light ray. 
Using this relation, we find 
\begin{eqnarray}
r_{1J}+\mbox{\boldmath $K$}\cdot\mbox{\boldmath $r$}_{1J}
&=&\frac{\theta^2 r_{1J}}{2}+O(\theta^4) , \\
\frac{r_{1J}+\mbox{\boldmath $K$}\cdot\mbox{\boldmath $r$}_{1J}}
{r_{2J}+\mbox{\boldmath $K$}\cdot\mbox{\boldmath $r$}_{2J}}
&=&1-\frac{2\mbox{\boldmath $n$}\cdot\mbox{\boldmath $B$}}
{r_{1J}\theta}+O\left(\frac{B^2}{r^2}\right) , 
\end{eqnarray}
where we introduced $r \sim r_{1J} \sim r_{2J}$. 
Using these approximations, Eq. $(\ref{eq:shapiro2})$ is rewritten as 
\begin{eqnarray}
\Delta(t_1,t_2)=-\frac{4GM}{c^3}&\Bigl(&
\frac{\mbox{\boldmath $n$}\cdot\mbox{\boldmath $B$}}
{r_{1J}\theta} \nonumber\\
&&+\frac{\mbox{\boldmath $B$}\cdot\mbox{\boldmath $v$}_J
-(\mbox{\boldmath $K$}\cdot\mbox{\boldmath $B$})
(\mbox{\boldmath $K$}\cdot\mbox{\boldmath $v$}_J)}
{c r_{1J}\theta^2}
+O(c^{-2}, c^{-1}B^2r^{-2}) \Bigr) , 
\label{eq:shapiro3}
\end{eqnarray}
which is in complete agreement with Eq. (12) of Kopeikin (2001). 
In deriving Eq. ($\ref{eq:shapiro3}$), we take account of 
the propagation only of light but not of gravity, since 
gravity propagation appears at $O(c^{-4})$. 
Hence, it turns out that the excess time delay given by 
Eq. $(\ref{eq:shapiro3})$ is due to nothing but the light-cone effect. 

Before closing this section, let us mention recent papers on this
issue: 
Clifford Will re-confirmed my conclusion 
by explicitly denoting the speed of light and gravity by different 
characters in his careful computations \cite{Will03}. 
Two more papers \cite{Faber,Samuel} support our conclusion. 
Nonetheless, Kopeikin wrote two papers 
\cite{Kopeikin02,Kopeikin03}: 
In one paper \cite{Kopeikin02}, he introduced an artificial time 
$\tau$ associated with the gravity speed $c_g$, and he claimed again 
that the excess depended on $V/c_g$, namely the gravity speed. 
However, we should notice a key that the velocity $V$ is $dx/d\tau$ 
but not the coordinate velocity $v=dx/dt$. There exits a relation 
$ct=c_g\tau$ in his computations, so that in the excess 
we can replace $V/c_g$ with $v/c$. 
The conclusion of the paper thus must be that 
the excess depends on the speed of light. 
In the other paper \cite{Kopeikin03}, when equations for the
gravitational field are integrated, he used the advanced Green 
function to show that the signature of the excess changed. 
This signature change apparently means the excess could be related 
with the gravity propagation. However, even for static case ($v=0$), 
signatures of two terms in a logarithmic function change 
in his expression of the Shapiro time delay, which must be unchanged. 
It's unbelievable! 
This happens because he integrated by mistake the null geodesics 
for light propagation on a future light cone emanating from the
observer but not a past light cone. 
Therefore, if the null geodesic were treated correctly on the past
light cone, the signature of the excess could not change for 
replacing the retarded Green function for gravitational fields 
with the advanced one. 
In short, this also is a proof that the excess has no relevance 
with the gravity propagation. 
By re-examining all of these recent papers, it seems that 
our conclusion has been well established. 

\section{Conclusion}
What was measured by the Jupiter event on the 8th of September 2002 
is not the gravity speed. 
Nonetheless, we should look on a positive side: The angular accuracy 
of their measurement by Very-Long Baseline Interferometry (VLBI) 
is a few tens micro-arcseconds, about one fifth of previous ones. 
This has great impacts on astrophysics. One is that a tighter 
constraint can be put on scalar-tensor theories of gravity through 
measuring the light deflection with the accuracy. Furthermore, 
it enables us to determine a parallax distance to stars at a few kpc, 
and to measure their proper motion by observations during several 
years. Therefore, the advanced VLBI must play an important role 
in cosmology as well as stellar and galactic physics.

\section*{Acknowledgments}
The author would like to thank L. Blanchet for his hospitality at the 
institute for Astrophysics at Paris, where this manuscript was written. 
This work was supported by a fellowship for a visiting scholar from 
the Ministry of Education of Japan.


\section*{References}
\begin{thebibliography}{99}
\bibitem{HKS}M. Homma, N. Kawaguchi and T. Sasao, Proc. SPIE, 
{\bf 4015}, 624 (2000).
\bibitem{KS}S. Kopeikin and G. Sch\"afer,  
\bibitem{Kopeikin01}S. Kopeikin, \Journal{Astrophys. J.}{556}{L1}{2001}
\Journal{Phys. Rev. D}{60}{124002}{1999} 
\bibitem{FK}E.B. Fomalont and S. Kopeikin, astro-ph/0302294.  
\bibitem{Asada}H. Asada, \Journal{Astrophys. J.}{574}{L69}{2002}
\bibitem{Will}C.M. Will, {\it Theory and Experiment in
Gravitational Physics} (Cambridge: Cambridge Univ. Press, 1993)
\bibitem{Thorne}K.S. Thorne, \Journal{Rev. Mod. Phys.}{52}{299}{1980}
\bibitem{PM}P.C. Peters and J. Mathews, 
\Journal{Phys. Rev.}{131}{435}{1963}
\bibitem{Taylor}J.H. Taylor, \Journal{Rev. Mod. Phys.}{66}{711}{1994}
\bibitem{Shapiro}I.I. Shapiro, \Journal{Phys. Rev. Lett.}{13}{789}{1964}
\bibitem{Will03}C.M. Will, astro-ph/0301145. 
\bibitem{Faber}J.A. Faber, astro-ph/0303346. 
\bibitem{Samuel}S. Samuel, astro-ph/0304006.
\bibitem{Kopeikin02}S. Kopeikin, gr-qc/0212121. 
\bibitem{Kopeikin03}S. Kopeikin, astro-ph/0302412. 
\end{thebibliography}
\end{document}